\documentclass[lettersize,journal]{IEEEtran}
\usepackage{amsmath,amsfonts}
\usepackage{algorithmic}
\usepackage{algorithm}
\usepackage{array}
\usepackage[caption=false,font=normalsize,labelfont=sf,textfont=sf]{subfig}
\usepackage{textcomp}
\usepackage{stfloats}
\usepackage{url}
\usepackage{verbatim}
\usepackage{graphicx}
\usepackage{multirow}
\usepackage{multicol}
\usepackage{booktabs}
\usepackage{colortbl}
\usepackage{color}
\usepackage{cite}
\begin{document}

\title{Transient Stability of Low-Inertia Power Systems with Inverter-Based Generation}

\author{Changjun He, Xiuqiang He,~\IEEEmembership{Member, IEEE}, and Hua Geng,~\IEEEmembership{Fellow, IEEE}, Huadong Sun,~\IEEEmembership{Senior Member, IEEE}, Shiyun Xu,~\IEEEmembership{Member, IEEE}
\thanks{This work was supported by the National Natural Science Foundation of China (U2166601, U2066602, and 52061635102). \textit{(Corresponding author: Hua Geng.)}

C. He, X. He, and H, Geng are with the Department of Automation, Beijing National Research Center for Information Science and Technology, Tsinghua University, Beijing 100084, China (e-mail: hcj20@mails.tsinghua.edu.cn; hxq19@tsinghua.org.cn; genghua@tsinghua.edu.cn).

Huadong Sun and Shiyun Xu are with the China Electric Power Research Institute (e-mail: sunhd6566@126.com; xushiyun@epri.sgcc.com.cn)

}
}



\maketitle

\begin{abstract}
This study examines the transient stability of low-inertia power systems with inverter-based generation (IBG) and proposes a sufficient stability criterion. In low-inertia grids, transient interactions are induced between the electromagnetic dynamics of the IBG and the electromechanical dynamics of the synchronous generator (SG) under a fault. For this, a hybrid IBG-SG system is established and a delta-power-frequency model is developed. Based on this model, new mechanisms of transient instability different from those of conventional power systems from the energy perspective are discovered. First, two loss-of-synchronization (LOS) types are identified based on the relative power imbalance owing to the mismatch between the inertia of the IBG and SG under a fault. Second, the relative angle and frequency will jump at the moment of a fault, thus affecting the system energy. Third, the cosine damping coefficient induces a positive energy dissipation, thereby contributing to the system stability. A unified criterion for identifying the two LOS types is proposed using the energy function method. This criterion is proved to be a sufficient stability condition for addressing the effects of the jumps and cosine damping coefficient on the system stability. The new mechanisms and effectiveness of the criterion are verified based on simulation results.
\end{abstract}


\section{Introduction}
\IEEEPARstart{L}{arge-scale} inverter-based generations (IBGs) replacing the typical synchronous generators (SGs) have been more connected to the power system in recent years. The main interfaces of IBGs to the grid are power electronic converters, which show more limited fault-tolerance capacity and less inertia than SGs \cite{ref24, shuai2018transient }. Moreover, multi-timescale dynamics are observed in IBG under a fault \cite{he2021modeling}. Thus, transient stability has become an essential issue that affects the stability of power grids. Previous studies are limited to the transient stability of IBG in a frequency-stiff grid scenario. Low-inertia power systems in which an unintended coupling is observed between the electromagnetic transients of the IBG and the electromechanical transients of the SG have received insignificant research attention.

In conventional power systems, transient stability (or transient angle stability) refers to the ability of SGs to maintain synchronization under a large disturbance. The dynamics of SGs are regulated by the physical inertial response, which is usually analyzed using a swing equation in single-machine-infinite-bus (SMIB) systems. Multimachines are divided into two clusters to achieve the equivalent two-machine system, which is further reduced into the equivalent SMIB system for analysis \cite{ref8, ref10, ref11, xue1989extended}. Studies on the transient stability of conventional power systems are mostly conducted in the electromechanical timescale.

In power systems with IBG, the definition of transient stability also considers whether a generation device (including the SG and the IBG) can maintain synchronization with other devices under a large disturbance \cite{ref13}. A grid-following phase-locked loop (PLL) is the most widely used synchronization strategy for inverters under a fault \cite{ref6}. Several studies have been conducted on the transient stability of PLL-based IBG in the electromagnetic timescale. A power-frequency model of PLL has been developed analogous to the swing equation of SGs \cite{ref19, ref24}. In other studies \cite{ref17, ref21, ref22, ref25}, the voltage-angle curve of the PLL is drawn in analogy to the power-angle curve of the SG. It has been reported that increasing and decreasing the proportional and integral gains, respectively, of the PLL can increase the damping ratio, thus improving the dynamic properties of the PLL \cite{ref16, ref17, ref18}. The voltage at the point of common coupling (PCC) varies considerably owing to the injection current of the IBG in weak grids. The interaction between the PCC voltage and PLL deteriorates the transient stability \cite{ref12, ref14, ref23, ref24, ref25}, and loss of synchronization (LOS) will occur because of the high grid impedance \cite{ref16}. In addition to the high grid impedance, the low-inertia property deteriorates the transient stability performance of weak grids \cite{ref17}. Subsequently, some studies have investigated transient stability in low-inertia grids with different types of generation devices \cite{ref12, ref13}. The transient stability of a hybrid system codominated by SGs and droop-controlled inverters outperforms that of the SG-based system \cite{ref12}. However, the most widely used grid-following devices are not employed in this hybrid system.

In summary, the understanding of the transient stability mechanism of low-inertia power systems with IBG in which the electromagnetic dynamics of the IBG and the electromechanical dynamics of the SG are coupled is lacking. To fill this gap, this study develops a hybrid IBG-SG system to describe the transient interactions between the IBG and SG in low-inertia grids. A delta-power-frequency model of the IBG-SG system is established to characterize the dynamic synchronization behaviors between the IBG and SG under the grid fault. Based on this model, the transient interactions between the IBG and SG based on active power relation are revealed. New mechanisms different from those of conventional power systems from the energy perspective are revealed. First, two LOS types, accelerating-type LOS and decelerating-type LOS, are observed based on the relative power imbalance of the IBG and SG under a fault. The mismatch between the inertia of the IBG and SG will exacerbate the imbalance, inducing the LOS under a fault. Second, the relative angle and frequency show abrupt jumps at the moment of a fault, thereby affecting the system energy. Third, a cosine damping coefficient is induced by the IBG. The varying damping term affords a positive energy dissipation and contributes to the system stability. A unified criterion for identifying the two LOS types is proposed using the energy function method, which offers advantages in terms of computation efficiency and intuitiveness. In addition, the conservativeness of this criterion is ensured under typical system parameters by considering the damping term and the jumps in the relative angle and frequency. The correctness of the analysis and effectiveness of the proposed criterion are verified based on simulation results.

The remainder of the paper is organized as follows. Section II introduces the modeling of the low-inertia power system with IBG and derives a delta-power-frequency model of the IBG-SG system. Section III analyzes the new transient stability mechanism from the energy perspective. Section IV presents the unified criterion for assessing the transient stability. Section V presents the simulations. Finally, Section VI presents the conclusions.

\section{System Modeling}
\begin{figure}[t]
\centering
\includegraphics[width=88mm]{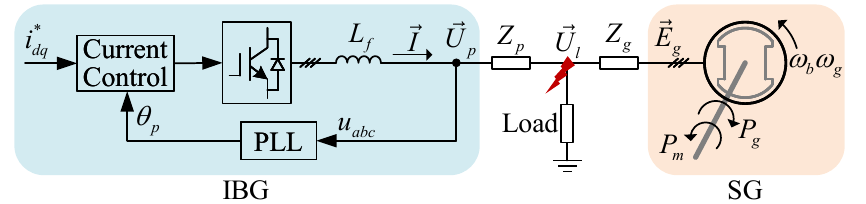}
\caption{Low-inertia power system with PLL-based IBG. IBG is equivalent to a PLL-synchronized current source under a fault. The grid side is equivalently represented using an SG. The IBG and SG are connected to the load through line impedances $Z_p$ and $Z_g$, respectively.}
\label{fig.1}
\end{figure}

To describe the transient behavior of the low-inertia power system with IBG, a hybrid IBG-SG system is established (Fig. \ref{fig.1}). For brevity, the following assumptions are proposed before modeling the dynamics of the IBG-SG system.
\begin{itemize}
\item{The current loop is ignored because it is considerably faster than the PLL \cite{ref16, ref18, ref27}.}
\item{The dynamics of the grid side are represented by the equivalent physical response of an SG.}
\item{The frequency deviation is usually small in power systems. Consequently, the line impedance is constant when ignoring the influence of the frequency on the reactance.}
\item{The load is characterized by a constant load impedance.}
\item{As a typical representation, assume that there is a three-phase symmetrical grounding fault occurring at the load bus.}
\end{itemize}

\begin{figure}[t]
\centering
\includegraphics[width=88mm]{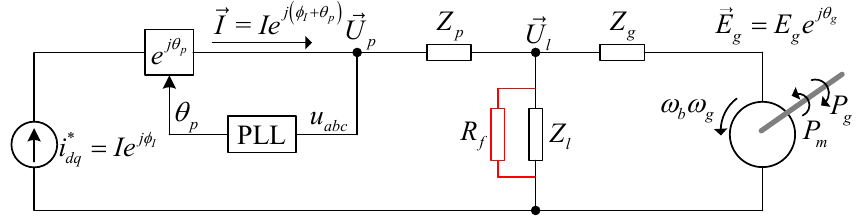}
\caption{Simplified circuit of the IBG-SG system.}
\label{fig.2}
\end{figure}

\begin{figure}[b]
\centering
\includegraphics[width=88mm]{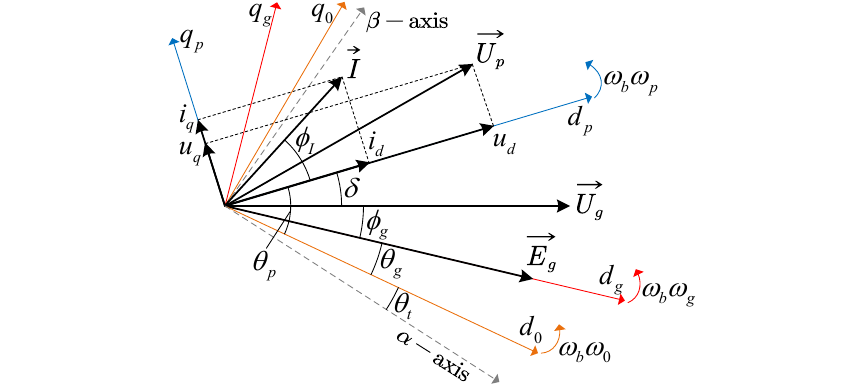}
\caption{Voltage and current vectors with different reference frames \cite{ref28}. The ${\alpha}$-, ${d_0}$-, ${d_g}$-, and ${d_p}$-axes represent the two-phase stationary reference frame, SRF, SG frame and PLL frame, respectively. ${\theta_t}=\omega_b \omega_0 t$ denotes the angle between the SRF and the two-phase stationary reference frame. ${\theta_g}$, the angle between the ${d_0}$- and ${d_g}$-axes, represents the SG frame angle in the SRF. ${\theta_p}$, the angle between the ${d_p}$- and ${d_0}$-axes, represents the PLL frame angle in the SRF. }
\label{fig.3}
\end{figure}

Based on these assumptions, the equivalent circuit of the IBG-SG system is developed (Fig. \ref{fig.2}). Here, ${R_f}$ represents the fault resistance, which is usually very small under a severe fault. Fig. \ref{fig.3} shows the voltage and current vectors with different reference frames. ${\omega_b}$ denotes the rotating speed of the synchronous reference frames (SRF) defined by the ${d_0}$- and ${q_0}$-axes, and ${\omega_0 = 1}$ pu represents its per unit value of speed. ${\omega_g}$ represents the rotating speed per unit value of the ${d_g}$-${q_g}$ frame of the SG. ${\omega_p}$ denotes the rotating speed per unit value of the ${d_p}$-${q_p}$ frame of the PLL. Then, the ${\alpha}$-, ${d_0}$-, ${d_g}$-, and ${d_p}$-axes represent the two-phase stationary reference frame, SRF, SG frame and PLL frame, respectively. ${\theta_t}=\omega_b \omega_0 t$ denotes the angle between the SRF and the two-phase stationary reference frame. ${\theta_g}$, the angle between the ${d_0}$- and ${d_g}$-axes, represents the SG frame angle in the SRF. ${\theta_p}$, the angle between the ${d_p}$- and ${d_0}$-axes, represents the PLL frame angle in the SRF. The definitions of angle ${\phi_g}$ and vector ${{\vec U}_g}$ are presented in \eqref{eq.3b}. The relative angle between the IBG and SG is defined as
\begin{align}
\label{eq.delta}
{\delta = \theta_p-\theta_g-\phi_g}.
\end{align}

\subsection{Swing Equation of SG}

The fault-on duration is usually short, and the primary frequency control of the SG is not triggered at this stage \cite{ref4}. Consequently, the dynamics of the SG are regulated using the physical inertial response. The dynamics of the SG can be expressed as a swing equation \eqref{eq.1}. The damping term of the SG is ignored because the damping effect of the SG is much weaker than that of the PLL \cite{ref22, ref16}.
\begin{align}
\label{eq.1}
\begin{aligned}
\frac{{d{\theta _g}}}{{dt}} & = {\omega _b}\left( {{\omega _g} - {\omega _0}} \right)\\
{T_g}\frac{{d{\omega _g}}}{{dt}} & = {P_m} - {P_g}
\end{aligned}
\end{align}
where ${T_g}$ denotes the inertia time constant, ${P_m}$ denotes the mechanical input power, and ${P_g}$ represents the electromagnetic output power. According to the circuit principle,
\begin{align}
\label{eq.2}
\begin{aligned}
{P_g} &= {\mathop{\rm Re}\nolimits} \left( {{{\vec E}_g}{{\vec I}_g}^*} \right)\\
&= \frac{{{E_g^2}}}{{\left| {{Z_g} + {Z_l}} \right|}}\cos {\phi _G} -{U_g}I\cos \left( {\delta + 2{\phi _g} + {\phi _I}} \right)
\end{aligned}
\end{align}
where
\begin{subequations}\label{eq.3}
\begin{align}
{\phi _G} &= \angle \left( {{Z_g} + {Z_l}} \right)\label{eq.3a}\\
{U_g}\angle {\phi _g} &= \frac{{{Z_l}}}{{{Z_g} + {Z_l}}}{E_g} \buildrel \Delta \over = {{\vec U}_g}\label{eq.3b}.
\end{align}
\end{subequations}

\subsection{Modeling of PLL-Based IBG}

\begin{figure}[b]
\centering
\includegraphics[width=88mm]{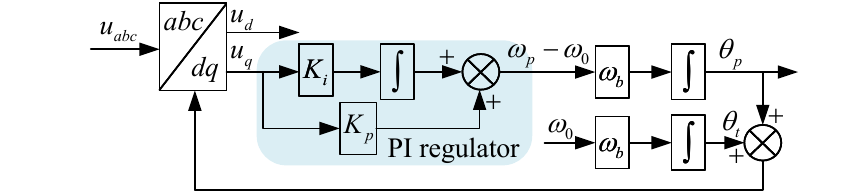}
\caption{Typical structure of the PLL. The structure is adapted so that the output of the PLL, the angle $\theta _p$, is referenced to the $d_0$-axis.}
\label{fig.4}
\end{figure}

The dynamics of the terminal filter and current loop of the IBG are usually neglected in transient stability analysis \cite{ref24}. Then, the IBG is deemed to be equivalent to a PLL-synchronized current source during the fault-on period. The currents ${{i_d} = i_d^*}$ and ${{i_q} = i_q^*}$ are guaranteed, where $i_{dq}^*=i_d^*+ji_q^*$ represents the current reference provided to the IBG. Based on the grid codes, the reactive current of IBG is required to be injected into the grid to support the voltage under a severe fault \cite{ref28}.

Fig. \ref{fig.4} shows a typical structure of the PLL. The PLL dynamics are expressed as
\begin{align}
\label{eq.4}
\begin{aligned}
\frac{{d{\theta _p}}}{{dt}} &= {\omega _b}\left( {{\omega _p} - {\omega _0}} \right)\\
\frac{{d{\omega _p}}}{{dt}} &= {K_p}\frac{{d{u_q}}}{{dt}} + {K_i}{u_q}
\end{aligned}
\end{align}
where $K_p$ and $K_i$ denote the proportional and integral gains of the PI regulator, respectively. The PLL detects the PCC voltage and estimates the ${q_p}$-axis component ${u_q}$ in the PLL frame. In the SRF, the PCC voltage is obtained as
\begin{align}
\label{eq.5}
\begin{aligned}
{{\vec U}_{p}} &= \frac{{{Z_l}}}{{{Z_g} + {Z_l}}}{E_g}\angle {\theta _g} + \left( {{Z_p} + {Z_g}\parallel {Z_l}} \right)I\angle \left( {{\theta _p} + {\phi _I}} \right)\\
 &= {U_g}\angle \left( {{\theta _g} + {\phi _g}} \right) + {Z_{eq}}I\angle \left( {{\theta _p} + {\phi _I} + {\phi _Z}} \right)
\end{aligned}
\end{align}
where
\begin{align}
\label{eq.6}
\begin{aligned}
{Z_{eq}}\angle {\phi _Z} &= {Z_p} + {Z_g}\parallel {Z_l}.
\end{aligned}
\end{align}

Subtracting ${\theta_p}$ from all the terms in \eqref{eq.5}, the PCC voltage vector is rewritten in the PLL frame as
\begin{align}
\label{eq.7}
{{\vec U}_{p}}\angle \left( { - {\theta _p}} \right){\rm{      }} = {U_g}\angle (-\delta) + {Z_{eq}}I\angle \left( {{\phi _I} + {\phi _Z}} \right).
\end{align}

Then, ${u_q}$ is expressed as
\begin{align}
\label{eq.8}
\begin{aligned}
{u_q} = - {U_g}\sin \delta + {Z_{eq}}I\sin \left( {{\phi _I} + {\phi _Z}} \right).
\end{aligned}
\end{align}

The reference power of the IBG is denoted as ${P_p^* = {u_d}i_d^*}$. It is only an auxiliary variable for analysis, and the actual reference provided to the inverter in engineering practices is the current $i_{dq}^*$. The actual power of IBG is ${{P_p} = {u_d}{i_d} + {u_q}{i_q}}$. Because ${{i_d} = i_d^*}$ and ${{i_q} = i_q^*}$ are given, the following equation can be obtained.
\begin{align}
\label{eq.9}
{u_q}{i_q} = {P_p} - P_p^*.
\end{align}

Based on \eqref{eq.4}--\eqref{eq.9}, the power-frequency relation \cite{ref19, ref24, ref29, ref30} of the PLL-based IBG is rewritten \eqref{eq.10}, resembling the swing equation of an SG.
\begin{align}
\label{eq.10}
\begin{aligned}
\frac{{d{\theta _p}}}{{dt}} &= {\omega _b}\left( {{\omega _p} - {\omega _0}} \right)\\
{T_p}\frac{{d{\omega _p}}}{{dt}} &= P_p^* - {P_p} - {D_p}\left( {{\omega _p} - {\omega _g}} \right)
\end{aligned}
\end{align}
where $T_p$ and $D_p$ represent the equivalent inertia time constant and equivalent damping coefficient of the IBG, respectively.
\begin{align}
\label{eq.11}
\begin{aligned}
{T_p} &= -\frac{I\sin {\phi _I}}{{{K_i}}}\\
P_p^* &=  {U_g}I\cos \delta \cos {\phi _I}  + {Z_{eq}}{I^2}\cos \left( {{\phi _I} + {\phi _Z}} \right) \cos {\phi _I}\\
{P_p} &= {U_g}I\cos \left( {\delta  + {\phi _I}} \right) + {Z_{eq}}{I^2}\cos {\phi _Z}\\
{D_p} &= -\frac{{{K_p}}}{{{K_i}}}\omega_b {U_g}I\sin {\phi _I} \cos \delta.
\end{aligned}
\end{align}

This model \eqref{eq.10} is only applicable to the case when $i_q\neq0$. And $i_q\neq0$ is easily guaranteed under faults because a reactive current must be injected into the grid to support the voltage according to the grid code \cite{ref24}. Equation \eqref{eq.10} shows that the imbalance between the reference and an actual powers of the IBG induces the frequency deviation of the PLL.

\subsection{Delta-Power-Frequency Model of the IBG--SG System}
Transient stability focuses on the relative movement between different generation devices. Thus, the relative frequency between the IBG and SG, defined as $\Delta\omega = {\omega _p} - {\omega _g}$, and the relative angle $\delta$ are two state variables of interest. When $\Delta\omega=0$, $\delta$ remains unchanged to maintain synchronization between the IBG and SG. When $\Delta\omega$ is unequal to 0 and $\delta$ increases or decreases to a large value during the fault-on period, the LOS occurs. Combining \eqref{eq.1} and \eqref{eq.10}, the IBG-SG system is reduced to an equivalent SMIB system \cite{ref10, xue1989extended}, which is modeled as
\begin{subequations}
\label{eq.12}
\begin{align}
\frac{{d\delta }}{{dt}} &= {\omega _b}\Delta \omega \label{eq.12a}\\
{T_{eq}}\frac{{d\Delta \omega }}{{dt}} &= {\Delta P_{in,eq}} - {\Delta P_{out,eq}} - {D_{eq}}\Delta \omega\label{eq.12b}
\end{align}
\end{subequations}
where $T_{eq}$ represents the system equivalent inertia time constant, ${\Delta P_{in,eq}}$ represents the system relative input power, ${\Delta P_{out,eq}}$ represents the system relative output power, and $D_{eq}$ represents the equivalent damping coefficient. They are expressed as follows:
\begin{subequations}\label{eq.13}
\begin{align}
{T_{eq}} =~&\frac{{{T_p}{T_g}}}{{{T_p} + {T_g}}}\label{eq:13a}\\
{\Delta P_{in,eq}} =~&\frac{{{T_g}}}{{{T_p} + {T_g}}}P_p^* - \frac{{{T_p}}}{{{T_p} + {T_g}}}{P_m}\label{eq:13b}\\
{\Delta P_{out,eq}} =~&\frac{{{T_g}}}{{{T_p} + {T_g}}}{P_p} - \frac{{{T_p}}}{{{T_p} + {T_g}}}{P_g}\label{eq:13c}\\
{D_{eq}} =~&\frac{{{T_g}}}{{{T_p} + {T_g}}}{D_p}\label{eq:13d}.
\end{align}
\end{subequations}

\begin{figure}[t]
\centering
\includegraphics[width=88mm]{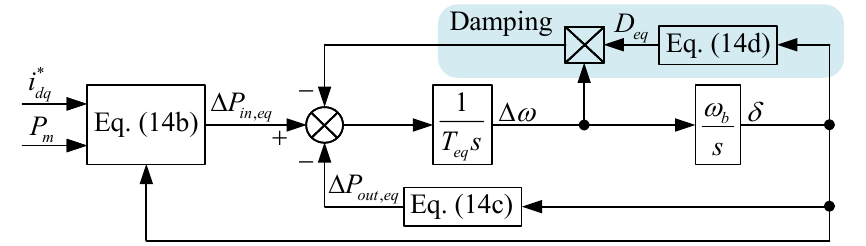}
\caption{Block diagram of the delta-power-frequency model \eqref{eq.12} of the IBG-SG system.}
\label{fig.5}
\end{figure}

\begin{figure}[t]
\centering
\includegraphics[width=88mm]{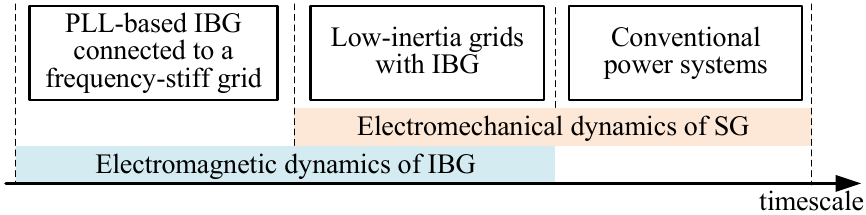}
\caption{Typical timescale in conventional power systems, PLL-based IBG connected to a frequency-stiff bus, and low-inertia grids with IBG.}
\label{time2}
\end{figure}

The imbalance between the relative input and output powers determines the dynamics of the relative frequency during the fault-on period; thus, the model \eqref{eq.12} is referred to as \emph{the delta-power-frequency model}. Fig. \ref{fig.5} shows the block diagram of the model \eqref{eq.12}. This figure indicates that the relative frequency is jointly governed by the input power/current of the SG/IBG and their dynamics. The transient stability of conventional power systems is only governed by the power dynamics of SGs; alternatively, the transient stability of IBG connected to a frequency-stiff grid is regulated only using the PLL-based power dynamics \cite{ref29}. Thus, a coupling is induced between the electromagnetic dynamics of the IBG and the electromechanical dynamics of the SG (Fig. \ref{time2}).

\section{Transient Stability Mechanism Analysis}
As analyzed above, the inertia of the IBG is governed by the control units. While the rotating part of the generator provides inertia to the SG. Accordingly, for analysis, \emph{the relative inertia ratio} of the SG to IBG is denoted as
\begin{align}
\label{eq.alpha}
\begin{aligned}
\alpha  = \frac{{{T_g}}}{{{T_p}}} = -\frac{{{K_i}{T_g}}}{I\sin {\phi _I}}.
\end{aligned}
\end{align}

Note that $\sin {\phi _I}\neq0$ is guaranteed during the fault-on period when the voltage dip is $<90\%$ \cite{ref16} because a certain amount of reactive current must be injected into the grid to support the voltage. Furthermore, there are no special requirements for the active current injection in the grid code. In particular, consider $i_d^* = 0$ and ${i_q^* = -I_r}$ during a severe fault \cite{ref15, ref18, ref24}, where $I_r$ represents the rated current of the IBG. Then, the current angle ${\phi _I}$ in Fig. \ref{fig.3} is ${90^ \circ }$. Subsequently, the following simplified formula of \eqref{eq.12b} is obtained from \eqref{eq.2}, \eqref{eq.3}, \eqref{eq.11}, \eqref{eq.13}, and \eqref{eq.alpha}. This form of mathematical expression is easier for analysis \cite{9662209}. Note that if ${\phi _I}$ is not ${90^ \circ }$, the same method can be applied to the following analysis.

\begin{align}
\label{eq.14}
\begin{aligned}
{T_{eq}}\frac{{d\Delta \omega }}{{dt}} = a + b\sin \left( {\delta + \varphi } \right) - d\cos \delta \Delta \omega
\end{aligned}
\end{align}
where $T_{eq}$, $a$, $b$, and $d\cos \delta $ are analogous to the inertia, constant input power, maximum electromagnetic power, and damping coefficient in the SG swing equation, respectively,

\begin{subequations}
\label{eq.15}
\begin{align}
a &= \frac{1}{{1 + \alpha }}\left( { - \alpha {Z_{eq}}{I^2}\cos {\phi _Z} + {\frac{{E_g^2\cos {\phi _G}}}{{\left| {{Z_g} + {Z_l}} \right|}}} - {P_m}} \right)\label{eq.15a}\\
b &= - \frac{{{U_g}I}}{{1 + \alpha }}\sqrt {{{\left( {\alpha + \cos 2{\phi _g} } \right)}^2} + {\sin ^2}2{\phi _g} } \label{eq.15b}\\
\varphi &= {\tan ^{ - 1}}\left( {\frac{{\sin  {2{\phi _g}} }}{{\alpha + \cos  {2{\phi _g}} }}} \right)\label{eq.15c}\\
d &= \frac{\alpha}{ {1 + \alpha } }\frac{K_p}{K_i}\omega_b{U_g}I\label{eq.15d}
\end{align}
\end{subequations}
and ${\tan ^{ - 1}}\left(  \cdot  \right)$ denotes the standard arctangent function with the range in the interval $[0,~\pi)$.

The energy function method offers advantages in terms of computational efficiency and good intuitiveness \cite{xue1989extended}; therefore, it is used to analyze the transient stability mechanisms of the IBG-SG system. The first integration method \cite{ref31} is used to obtain the energy function of the system as
\begin{align}
\label{eq.16}
\begin{aligned}
&{\rm{ }}V\left( {\delta ,{\rm{ }}\Delta \omega } \right)\\
 =~&{\frac{1}{2}{\omega _b}{T_{eq}}{{ {\Delta \omega } }^2}} { - a\delta  + b\cos \left( {\delta  + \varphi } \right)} + { d\int {\left( {\cos  \delta  \Delta \omega } \right)} d\delta }+ \lambda 
\end{aligned}
\end{align}
where $\lambda$ is an arbitrary constant, which depends on the reference potential energy surface.

\subsection{Two Types of New LOS Scenarios}
Ignoring the damping term, the energy function of the system is expressed as

\begin{align}
\label{eq.17}
\begin{aligned}
V\left( {\delta ,\Delta \omega } \right) = \underbrace {\frac{1}{2}{\omega _b}{T_{eq}}{\Delta {\omega ^2}}}_{{E_k}}\underbrace { - a\delta  + b\cos \left( {\delta  + \varphi } \right) + \lambda }_{{E_p}}.
\end{aligned}
\end{align}

\begin{figure}[h]
\centering
\includegraphics[width=88mm]{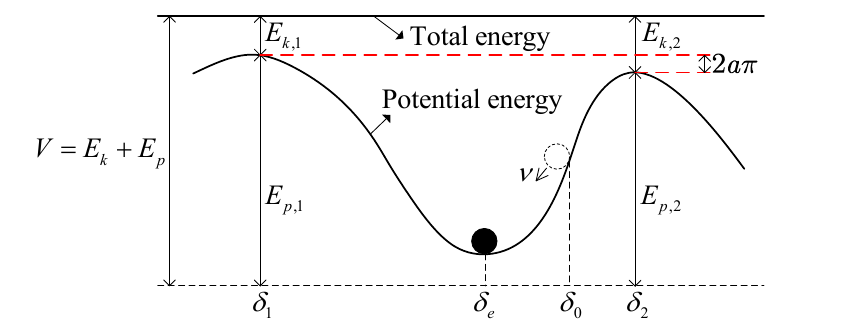}
\caption{Relation between different energies. At UEPs $\delta_{1}$ and $\delta_{2}$, the potential energy reaches the local maximum and the kinetic energy is the smallest. At SEPs $\delta_e$, the potential energy reaches the local minimum and the kinetic energy is the maximum.}
\label{fig.6-2}
\end{figure}

The energy function $V$ comprises kinetic energy ${E_k}$ and potential energy ${E_p}$, as shown in Fig. \ref{fig.6-2}. Previous studies have shown that the unstable equilibrium point (UEP) is critical in assessing the transient stability \cite{ref12,ref24}. Consequently, the movement in one typical cycle between two adjacent UEPs ($\delta_1$ and $\delta_2$) is studied here. The energies of these two UEPs are

\begin{align}
\label{eq.18}
\begin{aligned}
V\left( {{\delta _1},\Delta {\omega _1}} \right) = \underbrace{\frac{1}{2}{\omega _b}{T_{eq}}{{\left( {\Delta {\omega _1}}\right)}^2}}_{{E_{k,1}}}\underbrace { - a{\delta _1} + b\cos \left({{\delta _1} + \varphi } \right) + \lambda }_{{E_{p,1}}}
\end{aligned}
\end{align}

\begin{align}
\label{eq.19}
\begin{aligned}
V\left( {{\delta _2},\Delta {\omega _2}} \right) = \underbrace{\frac{1}{2}{\omega _b}{T_{eq}}{{\left( {\Delta {\omega _2}}\right)}^2}}_{{E_{k,2}}}\underbrace { - a{\delta _2} + b\cos \left( {{\delta _2} + \varphi } \right) + \lambda }_{{E_{p,2}}}.
\end{aligned}
\end{align}

\emph{The law of conservation of energy} states that the total energy remains unchanged during the fault-on period. Thus, we achieve

\begin{align}
\label{eq.20}
\begin{aligned}
V\left( {{\delta _2},\Delta {\omega _2}} \right) = V\left( {{\delta _1},\Delta {\omega _1}} \right).
\end{aligned}
\end{align}

Combining \eqref{eq.18}--\eqref{eq.20} yields a change in the kinetic energy

\begin{align}
\label{eq.21}
\begin{aligned}
{E_{k,2}} - {E_{k,1}} = 2a\pi.
\end{aligned}
\end{align}

\begin{table*}[b]
  \centering
  \caption{Influence of Parameters on Two Types of LOS}
    \begin{tabular}{ccccccc}
    \toprule
    \toprule
    Parameter & Description & Property & Power relation & Relative motion & Sign of $a$  & Type of LOS scenario \\
    \midrule
    \multirow{2}*{$\alpha$} & \multirow{2}*{Relative inertia ratio} & Large & $\alpha \Delta P_{p,s}<\Delta P_{g,s}$ & ${\omega _p}\searrow\searrow, ~{\omega _g}\searrow$ & $-$ & Decelerating-type LOS \\
    \cmidrule{3-7}
    {}&{}& Small & $\alpha \Delta P_{p,s}>\Delta P_{g,s}$ & ${\omega _p}\searrow, ~{\omega _g}\searrow\searrow$ & $+$ & Accelerating-type LOS \\
    \bottomrule
    \bottomrule
    \end{tabular}%
  \label{table0}%
\end{table*}%

\subsubsection{Accelerating-Type LOS}
${E_{k,2}}>{E_{k,1}}$ is obtained if $a>0$ from \eqref{eq.21}, implying that the kinetic energy increases during the movement from the left UEP $\delta_1$ to the right UEP $\delta_2$. $\delta$ continues to increase, and the PLL frame rotates faster than the SG frame, affording the accelerating-type LOS. However, when $\delta$ decreases from $\delta_2$ to $\delta_1$, the kinetic energy decreases and will finally become 0 because of its non-negative property. The system will therefore operate stably at one stable equilibrium point (SEP) under the damping effect. In summary, when $a>0$, only the accelerating-type LOS can occur but not the decelerating-type LOS.
\subsubsection{Decelerating-Type LOS}
Similarly, if $a<0$, ${E_{k,2}}<{E_{k,1}}$ is obtained. The kinetic energy increases during the movement from the right UEP $\delta_2$ to the left UEP $\delta_1$. $\delta$ continues to decrease, and the PLL frame rotates slower than the SG frame, affording the decelerating-type LOS. When $a<0$, the decelerating-type LOS instead of the accelerating-type LOS can occur.

\begin{figure}[t]
\centering
\includegraphics[width=88mm]{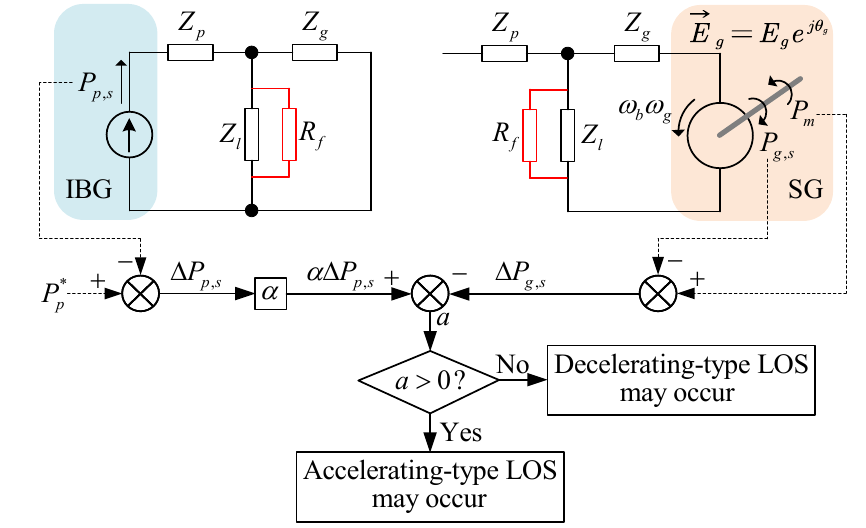}
\caption{Two possible types of LOS. The equivalent power shortages of IBG and SG during the fault-on period determine that type of LOS implying may occur.}
\label{fig.7}
\end{figure}

\emph{The accelerating-type LOS and the decelerating-type LOS may occur only when $a>0$ and $a<0$, respectively}. Based on \eqref{eq.15a}, $a$ is expressed as

\begin{align}
\label{eq.22}
\begin{aligned}
a =~&\frac{1}{{1 + \alpha }}\left( { - \alpha {Z_{eq}}{I^2}\cos {\phi _Z} + \frac{{E_g^2\cos {{\phi _G}} }}{{\left| {{Z_g} + {Z_l}} \right|}} - {P_m}} \right)\\
 =~&\frac{1}{{1 + \alpha }}\left( {  \underbrace {\alpha \left( {P_{p}^* - {P_{p,s}}} \right)}_{\alpha \Delta {P_{p,s}}} - \underbrace {\left( {P_m} - {{P_{g,s}}} \right)}_{\Delta {P_{g,s}}}} \right)
\end{aligned}
\end{align}
where $P_{p, s}$ and $\Delta {P_{p, s}}$ denote the output power and power shortage of the IBG, respectively, when the SG is removed under a fault (Fig. \ref{fig.7}). $P_{g, s}$ and $\Delta {P_{g, s}}$ denote the output power and power shortage of the SG, respectively, when removing the IBG under a fault. Note that the power of the IBG, $\Delta {P_{p, s}}$, is multiplied with $\alpha$ to obtain the equivalent power $\alpha\Delta {P_{p, s}}$; this is called \emph{the equivalent power shortage} of the IBG. If the equivalent power shortage of the IBG is larger than that of the SG (i.e., $a>0$), the frequency difference between the IBG and SG gradually increases and the accelerating-type LOS may occur. Alternatively, if the equivalent power shortage of the IBG is smaller than that of SG (i.e., $a<0$), the frequency difference between the IBG and SG gradually decreases and the decelerating-type LOS may occur.

The power relation between the IBG and SG considerably depends on the relative inertia ratio $\alpha$ (Table \ref{table0}). Moreover, the fault resistance will affect this power relation. Note that the SG will decelerate during the fault-on period with a low power penetration of the SG and a small external resistance owing to the fault. Thus, the findings suggest that a good match of the inertia between the IBG and SG balances their equivalent powers, thereby playing an important role in stabilizing the system under a fault.

\subsection{Effect of the Initial State Jump}
At the moment of the fault, the grid structure and parameters change, inducing a sudden change in the state variables of the system, i.e., $\delta$ and $\Delta \omega$. Based on the definitions of the state variables, the values at the moment of a fault are obtained as follows:
\begin{align}
\label{eq.34}
\begin{aligned}
{\delta _{0 + }} &= {\delta _{0 - }} + {\phi _g} - {{\phi}'_g}\\
\Delta {\omega _{0+}} &= {K_p}{u_{q,{0+}}}
\end{aligned}
\end{align}
where the subscripts ``$0-$" and ``$0+$" denote the prefault variable and the variable during fault, respectively; ${\phi _g}$ and ${{\phi}'_g}$ represent the angles during the prefault and during fault, respectively.

Under the effect of the initial state jumps, the kinetic and potential energies in \eqref{eq.17} will change to ${E_{k,0+}}$ and ${E_{p,0+}}$, respectively. In addition, the total energy of the system $V\left( {{\delta _{0 + }}, \Delta {\omega _{0 + }}} \right)$ is also altered:

\begin{align}
\label{eq.initial}
\begin{aligned}
~&V\left( {{\delta _{0 + }}, \Delta {\omega _{0 + }}} \right)\\
=~&\underbrace {\frac{1}{2}{\omega _b}{T_{eq}}{(\Delta {\omega_{0+}) ^2}}}_{{E_{k,0+}}}\underbrace { - a\delta_{0+}  + b\cos \left( {\delta_{0+}  + \varphi } \right) + \lambda }_{{E_{p,0+}}}.
\end{aligned}
\end{align}

\subsection{Effect of the Cosine Damping Coefficient}
The effect of the varying cosine damping coefficient has rarely been studied comprehensively. Based on the construction of a comprehensive energy function, the cosine damping term introduced by the IBG yields a positive energy dissipation. Consequently, the damping effect on the system stability is proved to be beneficial. The proof is shown below.

The LOS must occur without an SEP in \eqref{eq.12}. Considering a situation where an SEP exists, the inequality $\left| a \right| < \left| b \right|$ is given \cite{zi2020problems,ref16}. In addition, under the conditions of typical system parameters, $\alpha \gg 1$ and $\varphi \approx 0$ are obtained from \eqref{eq.15a} and \eqref{eq.15c}. We set the right term of \eqref{eq.14} to zero, and the relative angles of the SEP and UEP are approximated as follows:
\begin{align}
\label{eq.29}
\begin{aligned}
{\delta _e} & \approx {\sin ^{ - 1}}\left( -{{a \mathord{\left/
 {\vphantom {a b}} \right.
 \kern-\nulldelimiterspace} b}} \right) \in \left( { - {\pi \mathord{\left/
 {\vphantom {\pi  2}} \right.
 \kern-\nulldelimiterspace} 2},~{\rm{ }}{\pi  \mathord{\left/
 {\vphantom {\pi  2}} \right.
 \kern-\nulldelimiterspace} 2}} \right)\\
{\delta _{1}} & =  - \pi  - {\delta _e}\\
{\delta _{2}} & = \pi  - {\delta _e}.
\end{aligned}
\end{align}

Based on \eqref{eq.16}, the comprehensive system energy comprises the kinetic energy ${E_k}$, potential energy ${E_p}$, and energy dissipation ${\Delta E_{dis}}$ from the damping term.
\begin{align}
\label{eq.30}
\begin{aligned}
&{\rm{ }}V\left( {\delta ,{\rm{ }}\Delta \omega } \right)\\
 =~&\underbrace {\frac{1}{2}{\omega _b}{T_{eq}}{{ {\Delta \omega } }^2}}_{{E_k}}\underbrace { - a\delta  + b\cos \delta  + \lambda }_{{E_p}} + \underbrace { d\int {\left( {\cos  \delta  \Delta \omega } \right)} d\delta }_{\Delta {E_{dis}}}
\end{aligned}
\end{align}

\hspace*{\fill}\\

\noindent\textbf{Lemma 1} (The first mean value theorem for the integral\cite{apostol1991calculus}). \emph{Let $\phi ,\psi :[a,b] \to \Re$ be continuous in $[a,b]$. If the sign of $\psi$ never changes in $[a,b]$, there exists a number $c \in [a,b]$ such that}
\begin{align}
\label{mean value theorem}
\begin{aligned}
\int_a^b {\phi \left( x \right)\psi \left( x \right)} dx = \phi \left( c \right)\int_a^b {\psi \left( x \right)} dx.
\end{aligned}
\end{align}

Based on the theorem, the two LOS types identified in Section III-A are discussed separately.
\subsubsection{Accelerating-Type LOS} Moving from $\delta_e$ to $\delta_{2}$ yields
\begin{align}
\label{eq.31}
\begin{aligned}
&\Delta {E_{dis,2}}\\
 =~&d\int_{{\delta _e}}^{{\delta _{2}}} {\left( {\cos  \delta  \Delta \omega } \right)} d\delta \\
 =~&d\int_{{\delta _e}}^{{\pi  \mathord{\left/
 {\vphantom {\pi  2}} \right.
 \kern-\nulldelimiterspace} 2}} {\left( {\cos  \delta  \Delta \omega } \right)} d\delta  + d\int_{{\pi  \mathord{\left/
 {\vphantom {\pi  2}} \right.
 \kern-\nulldelimiterspace} 2}}^{{\delta _{2}}} {\left( {\cos  \delta  \Delta \omega } \right)} d\delta \\
 =~&d\Delta {\omega _ {a1}}\int_{{\delta _e}}^{{\pi  \mathord{\left/
 {\vphantom {\pi  2}} \right.
 \kern-\nulldelimiterspace} 2}} {\left( {\cos \delta } \right)} d\delta  - d\Delta {\omega _ {a2}}\int_{{\delta _2}}^{{\pi  \mathord{\left/
 {\vphantom {\pi  2}} \right.
 \kern-\nulldelimiterspace} 2}} {\left( {\cos \delta } \right)} d\delta \\
 =~&d\left( {1 - \sin {\delta _e}} \right)\left( {\Delta {\omega _ {a1}} - \Delta {\omega _ {a2}}} \right)
\end{aligned}
\end{align}
where $\Delta {\omega _ {a1}}$ and $\Delta {\omega _ {a2}}$ represent values within the deceleration interval $\left( {{\delta _e},~{\pi  \mathord{\left/ {\vphantom {\pi  2}} \right. \kern-\nulldelimiterspace} 2}} \right)$ and $\left( {{\pi  \mathord{\left/ {\vphantom {\pi  2}} \right. \kern-\nulldelimiterspace} 2},~{\delta _{2}}} \right)$, respectively. Because ${\delta _e} \in \left( { - {\pi  \mathord{\left/
 {\vphantom {\pi  2}} \right.
 \kern-\nulldelimiterspace} 2},{\pi  \mathord{\left/
 {\vphantom {\pi  2}} \right.
 \kern-\nulldelimiterspace} 2}} \right),{\delta _2} \in \left( {{\pi  \mathord{\left/
 {\vphantom {\pi  2}} \right.
 \kern-\nulldelimiterspace} 2},{{3\pi } \mathord{\left/
 {\vphantom {{3\pi } 2}} \right.
 \kern-\nulldelimiterspace} 2}} \right)$, the sign of $\cos \delta $ will not change in the internal $\left( {{\delta _e},{\pi  \mathord{\left/
 {\vphantom {\pi  2}} \right.
 \kern-\nulldelimiterspace} 2}} \right)$ and $\left( {{\pi  \mathord{\left/
 {\vphantom {\pi  2}} \right.
 \kern-\nulldelimiterspace} 2},{\delta _2}} \right)$. Thus, \emph{the first mean value theorem for the integral} can be used in \eqref{eq.31}. In the interval $\left( {{\delta _e},~{\delta _{2}}} \right)$, the representation of $\Delta {\omega _ {a2}} < \Delta {\omega _ {a1}}$ is drawn for ${{d\Delta \omega } \mathord{\left/
 {\vphantom {{d\Delta \omega } {dt}}} \right.
 \kern-\nulldelimiterspace} {dt}} < 0$. Then, $\Delta {E_{dis,2}} > 0$ is obtained from \eqref{eq.31}. In particular, if the maximum relative angle ${\delta _{max}}<\delta_{2}$, $d\int_{{\pi  \mathord{\left/ {\vphantom {\pi  2}} \right. \kern-\nulldelimiterspace} 2}}^{{\delta _{max}}} {( {\cos  \delta  \Delta \omega )} } d\delta>d\int_{{\pi  \mathord{\left/ {\vphantom {\pi  2}} \right. \kern-\nulldelimiterspace} 2}}^{{\delta _{2}}} {( {\cos  \delta  \Delta \omega )} } d\delta$ in \eqref{eq.31} is obtained. Consequently, the conclusion $\Delta {E_{dis,2}} > 0$ still holds. The analysis and conclusion are also applicable to the backward movement from $\delta_2$ to the SEP $\delta_{e}$.

\subsubsection{Decelerating-Type LOS}
Analogous to \eqref{eq.31}, the energy dissipation of the damping term from $\delta_e$ to $\delta_{1}$ is derived as
\begin{align}
\label{eq.32}
\Delta {E_{dis,1}}=d\left( {1 + \sin {\delta _e}} \right)\left( {\Delta {\omega _{d2}} - \Delta {\omega _{d1}}} \right)
\end{align}
where $\Delta {\omega _{d1}}$ and $\Delta {\omega _{d2}}$ denote values within the acceleration interval $\left( {{\delta _1},-{\pi  \mathord{\left/ {\vphantom {\pi  2}} \right. \kern-\nulldelimiterspace} 2}} \right)$ and $\left( -{{\pi  \mathord{\left/ {\vphantom {\pi  2}} \right. \kern-\nulldelimiterspace} 2},~{\delta _{e}}} \right)$, respectively. In the acceleration interval $\left( {{\delta _1},~{\delta _{e}}} \right)$, the representation of $\Delta {\omega _{d2}} > \Delta {\omega _{d1}}$ is drawn for ${{d\Delta \omega } \mathord{\left/
 {\vphantom {{d\Delta \omega } {dt}}} \right.
 \kern-\nulldelimiterspace} {dt}} > 0$. Then, $\Delta {E_{dis,1}} > 0$ is obtained from \eqref{eq.32}. In particular, if the minimum relative angle is larger than $\delta_{1}$, $\Delta {E_{dis,1}} > 0$ still holds. The analysis and conclusion are also applicable to the forward movement from $\delta_1$ to the SEP $\delta_{e}$.
 
In summary, $\Delta {E_{dis}} > 0$ always holds for all one-direction movements from or to the SEP or their combinations, irrespective of the trajectory. In such cases, \emph{the varying cosine damping term introduces a positive energy dissipation in the system with typical parameters ($\alpha \gg 1$, $\varphi \approx 0$)}, benefiting the transient stability of the IBG-SG system.

\section{Transient Stability Assessment}
Two possible types of LOS scenarios are identified in Section III-A. When an SEP exists, the occurrence of the LOS will depend on the initial state and dynamic performance of the system \cite{ref16, ref17}.

\subsection{Unified Criterion for Two Types of New LOS Scenario}

\begin{figure}[h]
\centering
\includegraphics[width=88mm]{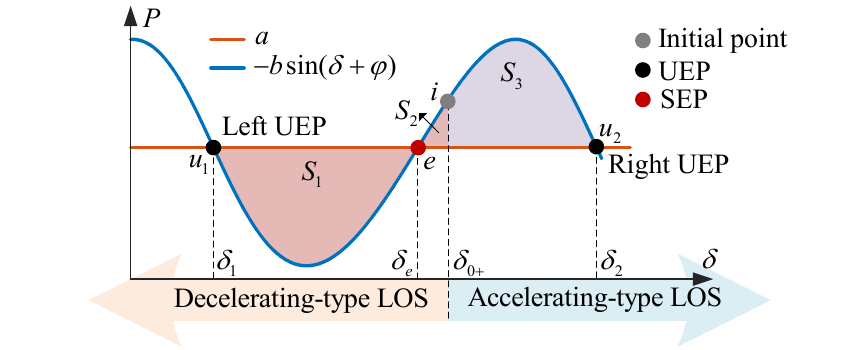}
\caption{$P$-$\delta$ curves of the delta-power-frequency model \eqref{eq.12}. In the interval $(\delta_{1},~\delta_e)$, $\Delta \omega$ increases for $a >- b\sin \left( {\delta + \varphi } \right) $. $S_1$ represents the acceleration area. Similarly, in the interval $(\delta_e,~\delta_{2})$, $\Delta \omega$ decreases for $a <- b\sin \left( {\delta + \varphi } \right) $. $S_2$ and $S_3$ represent the deceleration areas.}
\label{fig.8}
\end{figure}
As shown in Fig. \ref{fig.8}, without considering the damping effect, the offset term $a$ and sine term $- b\sin \left( {\delta + \varphi } \right)$ determine the movement. The motion from the initial point $i(\delta_0+,\Delta \omega_0+)$ to a point $M(\delta_M,\Delta \omega_M)$ ($M$ can be any point, such as point $u_1$ or $u_2$) satisfies
\begin{align}
\label{DELTA}
\begin{aligned}
{E_{k,0 + }} - {E_{k,M}} = \underbrace {{E_{p,M}} - {E_{p,0 + }}}_{\Delta {E_{p,M}}}
\end{aligned}
\end{align}
where ${\Delta {E_{p,M}}}$ denotes the change in potential energy from the initial point $i$ to the point $M$. The critical condition to reach the point $M$ is $E_{k,M}=0$. In other words, the critical initial kinetic energy ${E_{k,0+(cri) }}$ is equal to the change in potential energy.
\begin{align}
\label{critical}
\begin{aligned}
{E_{k,0+(cri) }} = {\Delta {E_{p,M}}}.
\end{aligned}
\end{align}

If the initial kinetic energy is smaller than the critical initial kinetic energy (i.e., ${E_{k,0+}}<{\Delta {E_{p,M}}}$), the system cannot reach the point $M$.

Based on the definition of \emph{the Riemann integral}, the area enclosed by the two curves $a$ and $- b\sin \left( {\delta + \varphi } \right)$ in Fig. \ref{fig.8} is equal to the change in potential energy. Then, the change in the potential energy from the initial point $\delta_e$ to UEPs $u_1$ and $u_2$ is represented as $\Delta {E_{p,1}}$ and $\Delta {E_{p,2}}$, respectively.
\begin{align}
\label{Ep}
\begin{aligned}
\Delta {E_{p,1}} = {S_1} - {S_2},\Delta {E_{p,2}} = {S_3}.
\end{aligned}
\end{align}

\subsubsection{Accelerating-Type LOS} 
If $a>0$ and the system reaches the right UEP $u_2$, the accelerating-type LOS will occur. Therefore, the stability condition for the accelerating-type LOS is given by
\begin{align}
\label{eq.24}
\begin{aligned}
{E_{k,0+}} < {S_3}.
\end{aligned}
\end{align}

\subsubsection{Decelerating-Type LOS}
If $a<0$ and the system reaches the left UEP $u_2$, the decelerating-type LOS will occur. Then, the stability condition for the decelerating-type LOS is reduced to
\begin{align}
\label{eq.25}
\begin{aligned}
{E_{k,0+}} < {S_1} - {S_2}.
\end{aligned}
\end{align}
\subsubsection{Unified Transient Stability Criterion} 
The system shows transient stability if neither the accelerating-type LOS nor the decelerating-type LOS occurs. Consequently, the stability conditions of \eqref{eq.24} and \eqref{eq.25} should be satisfied. Then, a unified criterion for the two LOS types is summarized: \emph{the system is transient stable during the fault-on period if \eqref{eq.28} is satisfied.}
\begin{align}
\label{eq.28}
\begin{aligned}
{E_{k,0+}} < \min \left\{ {{S_1} - {S_2},{\rm{ }}{S_3}} \right\}.
\end{aligned}
\end{align}

\begin{figure}[h]
\centering
\includegraphics[width=88mm]{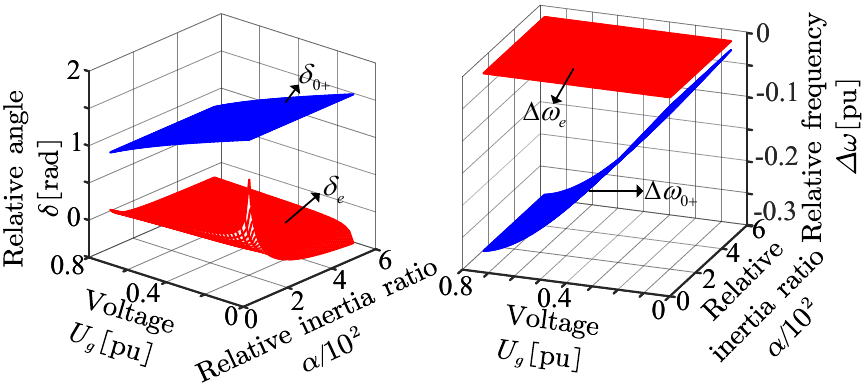}
\caption{Curves of the initial state and SEP state variables under different relative inertia ratios and fault depths. The blue surface denotes the initial value at the moment of the fault, and the red surface denotes the SEP value.}
\label{3Dv3}
\end{figure}

\subsection{Conservativeness Analysis on the Criterion}
In the SG swing equation, a constant positive damping coefficient increases the acceleration area and reduces the deceleration area, affording conservativeness of the criterion without considering the damping term \cite{ref31,ref1,zi2020problems}. Alternatively, a constant negative damping coefficient affords a radical result. However, the effect of the varying damping coefficient on the transient stability is not known comprehensively in the previous studies. In the single-converter-infinite-bus system, some studies have imposed additional restrictions on the range $(\delta_{\min},~\delta_{\max})$ to ensure a positive damping coefficient \cite{ref20, ref22, ref24}. In this section, we demonstrate that no additional restrictions are required to guarantee the conservativeness of the criterion under the conditions of typical system parameters.

To ensure the good dynamic performance of the PLL \cite{ref24} and existence of SEP during the fault-on period, the PLL integral gain $K_i$ and SG inertia $T_g$ are adopted in a certain range: $K_iT_g$ is roughly $36$--$415$. Fig. \ref{3Dv3} shows the curves of the initial state and SEP state variables under these typical parameters. The blue surface denotes the initial values at the moment of the fault, and the red surface denotes SEP values. This figure shows that $\delta _{0 + } > \delta _e$ and $\Delta \omega _{0 + } < \Delta \omega _e$ always hold under the conditions of the aforementioned system parameters. As shown in Fig. \ref{fig.8}, starting from the initial point $i$, $\delta$ decreases ($\Delta \omega<0$) faster (${{d\Delta \omega } \mathord{\left/ {\vphantom {{d\Delta \omega } {dt}}} \right. \kern-\nulldelimiterspace} {dt}}<0$) to the SEP $e$. Based on the conclusion presented in Section III-C, a positive damping dissipation is achieved in the one-direction movements from or to the SEP. Consequently, the movement from the initial point to the UEP causes the energy dissipation of the system, i.e., $\Delta {E_{dis}} > 0$.

A positive energy dissipation ${\Delta E_{dis}}$ yields a reduced kinetic energy ${E_k}$, indicating that the damping term contributes to the system stability. If the criterion shows that the typical IBG-SG system is stable, the system will operate stably. This indicates that the criterion is a sufficient condition and exhibits strict conservativeness for stability assessment under the aforementioned conditions of typical parameters.        



\section{Simulation Verification}
To adequately verify the correctness of the analysis and effectiveness of the stability criterion, simulations of the stable/unstable conditions should be conducted on the IBG-SG system in Fig. \ref{fig.1}. Simulation experiments should be performed under the same conditions as the analytical model \eqref{eq.12}. As a result, a phasor model of Fig. \ref{fig.1} is developed in SIMULINK, where the IBG interface is represented by a current source \cite{IEC}. Table \ref{table1} presents the parameters based on the literature \cite{zi2020problems,ref16}. A fault occurs at $t$ = 0.5~s. The time period until 3~s is shown to exhibit the destabilization/stabilization process of the system.

\begin{table}[t]
\centering
\caption{System Parameters}
\label{table1}
\begin{tabular}{ c c c }
\toprule
\toprule
Symbol & Description & Value\\
\cmidrule{1-3}
$S_b$ & Base capacity & 20 kVA\\
$U_b$ & Base voltage & 690 V\\
$\omega_b$ & Base frequency & $100\pi$ rad/s\\
$E_g$ & Internal electric potential of SG & 1.05 pu\\
$P_m$ & Mechanical power of SG & 0.2465 pu\\
$I_r$ & Rated current of IBG & 0.8 pu\\
$Z_l$ & Load impedance & 0.99 + j0.1 pu\\
$Z_g$ & Line impedance on the SG side & 0.01 + j0.1 pu\\
$Z_p$ & Line impedance on the IBG side & 0.01 + j0.3 pu\\
\bottomrule
\bottomrule
\end{tabular}
\end{table}

\begin{figure*}[b]
\centering
\includegraphics[width=180mm]{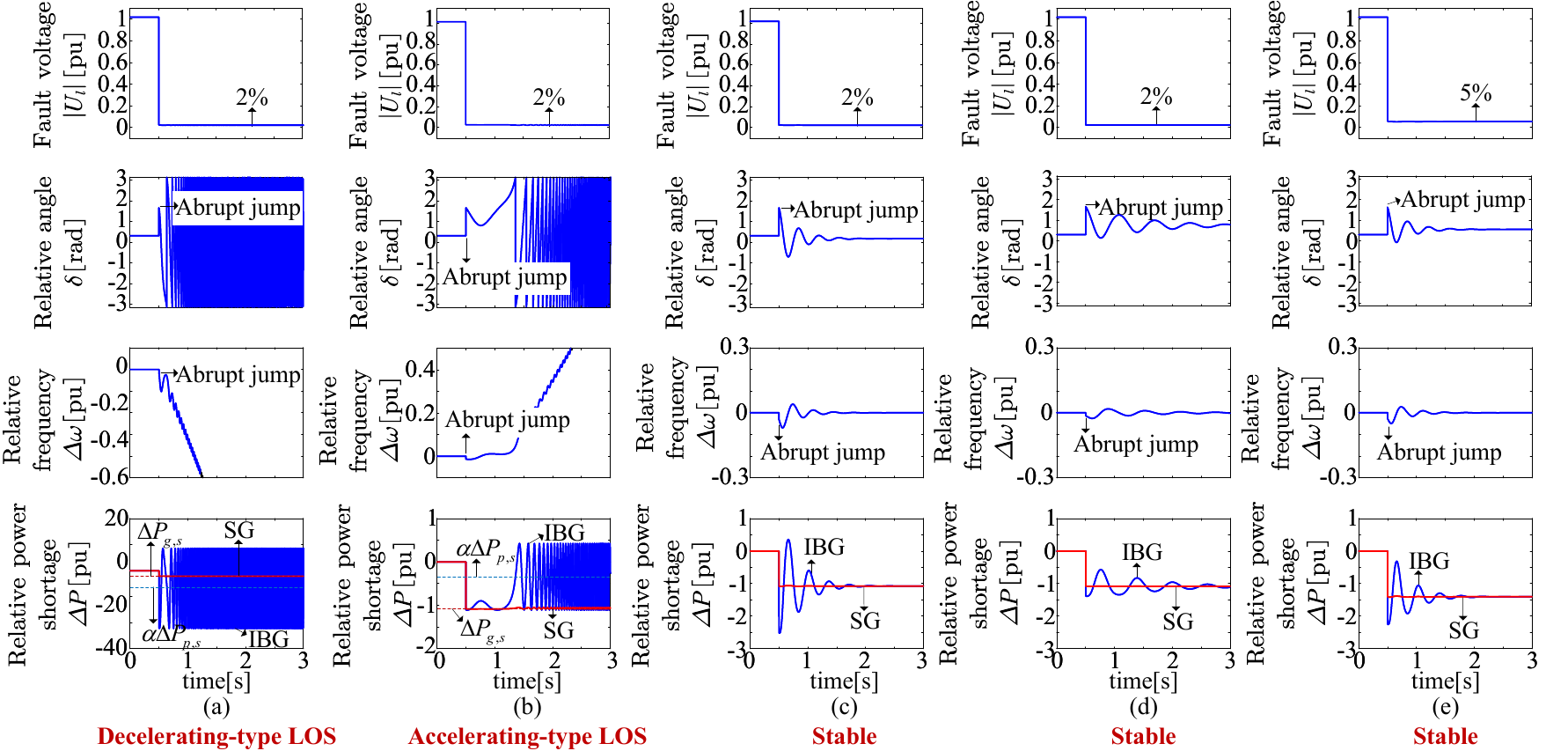}
\caption{Simulation results of the two types of LOS. (a) Case I: decelerating-type LOS occurs when $\alpha \Delta {P_{p, s}}< \Delta P_{g, s}$. (b) Case II: accelerating-type LOS occurs when $\alpha \Delta {P_{p, s}}> \Delta P_{g, s}$. (c) Case III: system becomes stable when $K_i$ is adjusted to 50 ${\rm{s}}^{-1}$, and $K_p=0.02K_i$ is the same in Cases II and III to keep the damping effect of the PLL. (d) Case IV: system becomes stable when $T_g$ is increased to 1 ${\rm{s}}$. (e) Case V: system becomes stable when the fault is less severe, $R_f$ is increased to 0.12 $\Omega$, and $|U_l|$ is increased to 5\%.}
\label{fig.12v2}
\end{figure*}

\subsection{Two Types of New LOS Scenarios}
The jumps in $\delta$ and $\Delta \omega$ at the moment of the fault are observed in all cases in Fig. \ref{fig.12v2}. The two types of new LOS scenarios are observed in Cases I and II (Fig. \ref{fig.12v2}). In Case I, a large $K_i$ affords a large $\alpha$, inducing a smaller equivalent power shortage of the IBG than that of the SG. $\Delta \omega$ decreases, and $\delta$ oscillates. Finally, decelerating-type LOS occurs. In Case II, a small $K_i$ affords a small $\alpha$. The equivalent power shortage of the IBG is larger than that of the SG, thus increasing $\Delta \omega$ and oscillating $\delta$.

\subsection{Influence of Parameters}
By adjusting the parameters considered in Case II, the system becomes stable in Cases III--V (Fig. \ref{fig.12v2} (b)--(d)). In Case II, the equivalent power shortage of the IBG is larger than that of the SG. $K_i$ of the PLL increases in Case III to yield an increased $\alpha$, thus changing the extreme imbalance between the equivalent power shortages of the IBG and SG. $\delta$ does not exceed left and right UEPs; thus, the system remains stable. In Case IV, $\alpha$ increases as the inertia of the SG increases, resulting in a stable system. An intuitive explanation for the stability phenomenon in Case IV is that the frequency of the SG changes at a slow pace when the inertia is large, and the IBG can keep up with the changes. Moreover, the fault becomes less severe in Case V. The redistribution of the power mitigates the equivalent power shortage of the SG and alleviates the severe mismatch between the power shortages of the IBG and SG to stabilize the system.

Table \ref{table2} shows that to enhance the transient stability of the system, $K_i$ should be adaptively selected to adjust the inertia of IBG to meet different operation conditions.

\subsection{Effectiveness of the Criterion}
Table \ref{table2} shows the results of the criterion assessment and simulation. The criterion results are consistent with the simulation results. The stability/instability of the system and the LOS type are effectively evaluated. The criterion is proved to be an effective method for the transient stability assessment of the IBG-SG system.

\section{Conclusion}
This study analytically investigated the transient stability of low-inertia power systems with IBG. A hybrid IBG-SG system is established to characterize the transient interactions between the electromagnetic dynamics of the IBG and the electromechanical dynamics of the SG in the low-inertia grid under a fault. To describe the transient behavior of the IBG-SG system, a delta-power-frequency model is developed. Based on the model, new mechanisms from the energy perspective are revealed. First, two new LOS types, accelerating-type LOS and decelerating-type LOS, are clarified based on the relative power imbalance between the IBG and SG. In addition, a good match between the inertia of the IBG and SG balances the powers and stabilizes the system. Second, $\delta$ and $\Delta \omega$ jump abruptly at the moment of a fault, affecting the system energy. Third, the cosine damping coefficient is induced by the IBG. The damping term introduces a positive energy dissipation, beneficially affecting the transient stability. A unified criterion for the two LOS types is adopted to evaluate the system stability using the energy function method. The conservativeness of the criterion is ensured when addressing the effects of the jumps and damping term on the transient stability. The accuracy of the analysis and the effectiveness of the criterion can be verified based on the simulation results.

Regarding the transient stability issue, the findings of this study provide enlightening significance for the PLL parameters tuning, relative inertia ratio shaping, development of grid codes for fault ride-through, and future grid design. Note that the frequency regulation should be added to ensure the frequency stability of the system, which is not discussed here. Other dynamic loops, such as the DC voltage control loop, and the stability during the fault recovery period are beyond the scope of this paper, which will be analyzed in our future work.

\begin{table*}[t]
\begin{center}
  \centering
  \caption{Results of the Criterion and Simulation}
  \label{table2}
    \begin{tabular}{ccccccccc}
    \toprule
    \toprule
    {Case} & $K_i$ & {$K_p$} & $T_g$ & $R_f$ & $|U_l|$ & {Condition} & {Criterion result} & {Simulation result} \\
    {}& [${{\rm{s}}^{ - 1}}$]& & [$\rm{s}$] & [$\Omega$] & [$\rm{pu}$] &{}&{}&{}\\
    \midrule
     I  & 220 & 1  & 0.8   & 0.05 & 0.02 & $a<0,~{E_{k,0+}} > {S_1} - {S_2}$ & Decelerating-type LOS & Decelerating-type LOS \\
    II   & 22  & 0.44 & 0.8 & 0.05 & 0.02 & $a>0,~{E_{k,0+}} > {S_3}$ & Accelerating-type LOS & Accelerating-type LOS \\
     III    & 50& 1  & 0.8 & 0.05 & 0.02 & $a>0,~{E_{k,0+}} < {S_3}$ & Stable & Stable \\
    IV   & 22 & 0.44   & 1 & 0.05 & 0.02 & $a>0,~{E_{k,0+}} < {S_3}$ & Stable & Stable \\
     V    & 22  & 0.44  & 0.8 & 0.12 & 0.05 & $a>0,~{E_{k,0+}} < {S_3}$ & Stable & Stable \\
    \bottomrule
    \bottomrule
    \end{tabular}%
\end{center}
\end{table*}

\bibliography{ref.bib}

\end{document}